\documentstyle[preprint,aps,prl]{revtex}
\tightenlines

\begin{document}
\draft
\preprint{To appear in Phys.Rev.E}
\title{Theory on  quench-induced pattern formation:  Application
to the isotropic to smectic-A phase transitions }

\author{Zhou    Haijun$^{1\;*}$ and   Ou-Yang      Zhong-can$^{1,2}$}
\address{
$^{1}$Institute of Theoretical Physics, Chinese Academy of Sciences,\\
P.O. Box 2735, Beijing 100080, China\\
$^{2}$Center for Advanced Study, Tsinghua University, Beijing 100084, China}
\date{August 2, 1998}

\maketitle
\begin{abstract}
 During catastrophic processes of environmental variations
of a thermodynamic system,
 such as rapid temperature decreasing, 
 many novel and complex patterns often form. 
 To understand such phenomena, a general mechanism is proposed based 
on the competition between heat transfer  and  conversion
of heat to other energy forms. We apply it to the smectic-A filament growth
process during quench-induced isotropic to smectic-A phase transition. 
Analytical forms for the buckling patterns are  derived
 and we find good agreement with experimental observation [Phys. Rev. {\bf E55} (1997) 1655].  
The present work strongly indicates that rapid cooling will lead to structural
transitions in the smectic-A filament at the molecular level 
to optimize heat conversion. The force associated with this
pattern formation process is estimated to be in the order of
$10^{-1}$ piconewton.
 \end{abstract}

\narrowtext

When a thermodynamic system in some initial stable phase is subjected to
rapid or abrupt environmental variations such as decreased temperature,
varied ionic conditions, etc., many novel patterns will often form. 
 For example, in
volcanic eruptions streams of lava are spurted out and cooled  rapidly,
during which many tensile bubbles are spontaneously formed  and the
lava becomes porous material [1, 2]. Similar bubbles and strips are 
 seen in supercooled biomembranes [3].  And 
 in  rapid growth process of carbon nanotubes
  initial straight tubes are observed to change into
regular helical shapes, even at fixed
environmental conditions [4].
 Other puzzling phenomena include
the kink instabilities of short DNA rings in solution: on adding some specific
ions, a circular ring  changes abruptly into polygonal ones [5].

All these transitions and pattern formations are radically different from
those slow, quasi-static processes discussed in conventional
 textbooks, such as the isothermal solid-liquid
transitions. In the
solidification process of water, for example,
 the transition takes place very slowly and
at any time during this process,  solid and liquid phases can be regarded
as in equilibrium with each other and with the environment, the bulk
temperature is kept constant and latent heat of transition is released.
However, for those catastrophic processes mentioned above [1-5], the quasi-static
and quasi-equilibrium assumptions are no longer satisfied since the system is
now far away from equilibrium and changes rapidly over time [6].
 Therefore, we have to create new ways and
new insights to properly understand them. 
 A deep
understanding of such phenomena   is needed both for its own
sake and for various practical purposes, such as the optimal growth conditions
for some important materials.
In recent years many research efforts have been devoted to this field [6-10].
For example, in Refs.~[6] and [7] (and in other references)
 the spinodal decomposition and formation of separate patterns
  in a binary mixture is studied both 
  theoretically and experimentally, using respectively
linear stability analysis and x-ray scattering technique.
And Refs.~[8-10] and [6] have investigated the solid-liquid interphase patterns 
during directional solidification processes induced by large temperature
gradient. A very important aspect of the mechanism for
 these pattern formation phenomena  is that the pattern formation
 is controlled mainly by the relatively slow diffusive transport process
 of heat energy
   out of the system [6].
Thus, It seems that this mechanism can not give a  satisfactory explanation to 
 those pattern formation processes  mentioned in the
first paragraph of this paper. This is partly because of the fact
that in these systems the effects of other
 energy forms become comparable to that of   heat energy [1-5], and they are
 anticipated to take a significant role
 in determining the formed patterns.
 
 In this paper,  we try to understand the  nature of these rapid pattern
formation processes by regarding them as quench-like ones induced by,
say, temperature decreasing. Our main reasoning is as follows. In a quench
process caused by rapid cooling, the thermal energy or heat of the system is
required to be evacuated as quickly as possible. However, this demand can not
be met just only through the relatively slow heat exchange process occurred
at the system surface, and a great part of the thermal energy are thus
anticipated to be converted into other energy forms [11, 12]: mechanical, elastical,
and maybe even {\it electromagnetical} ones, which can accommodate large
amount of energy quickly and efficiently. Therefore, during the quench
process, because of the competition between
heat transfer and heat conversion,
 the structures of the system, both  at   macroscopic level and at
 molecular level, might change dramatically, and many novel
and complex patterns are anticipated to form spontaneously. It resembles in
some sense the rapid expansion of ideal gases, which is regarded as an
adiabatic process since the system has no time to exchange heat with the
environment.

To check the validity of this perspective and show how to
implement it in actual practice, we  investigate into a specific
phenomenon based on this insight, namely, 
the growth and pattern formation process of smectic-A
(Sm-A) filaments from an isotropic (I) liquid phase induced by rapid
cooling [13]. Although the subject is quite specific, the general reasoning is the
same and can be applied to many other pattern formation processes. Our
theoretical results show that: On cooling, an initial straight filament will buckle
and take on some specific kind of curved configurations which can be well
described by elliptic integrals. Good agreements with the
experiment of Ref.~[14] have been attained, indicating the usefulness and
robustness of our treatment.  As also speculated in Ref.~[14], the present work
confirms that structural transitions in the Sm-A filament at the molecular level
are required  in order to optimize heat conversion. The force associated 
with this pattern formation is also discussed in this paper.

As observed in the I--Sm-A phase transition process [14], at the initial stage,
thin, hollow, and straight Sm-A liquid-crystalline tubes appear from the I
 phase on cooling. As the temperature is further decreased, these
straight filaments buckle at the growing ends and take on  somewhat complex
serpentine forms, the amplitudes of which become more and more large. The
filaments are metastable and eventually transform to compact domains after
the filaments become extremely convoluted. Although this 
transition phenomenon seems to be very complex, we 
found  its physics is actually  simple and
the formed patterns can be well described by analytical functions.

As we have known, the net difference in the energy between a thin Sm-A
filament and the I phase is the sum of the following three terms [13, 14]:
 
 (1) the volume
free energy change due to I--Sm-A transition $F_V=-g_0 V=-\pi (\rho _o^2
-\rho_i^2)\int g_0 d s$, where $g_0$ is the difference in the Gibbs free energy
densities between I and Sm-A phases and $V$ is the volume of the Sm-A
nucleus, and $\rho_o$ and $\rho_i$ are respectively the outer and inner
radii of the Sm-A microtube, $s$ is the arc length along the central 
line ${\bf r}$
 of it; 
 
 (2) the surface energy of the outer and inner Sm-A--I
interfaces $F_A=\gamma (A_o+A_i)=2\pi (\rho_o+\rho_i)\gamma \int d s$,
where $\gamma $ is the Sm-A--I interfacial tension, and $A_o$ and $A_i$ are
the surface areas of the outer and inner surfaces, respectively; 

(3) the
curvature elastic energy of the Sm-A filament $F_c=(k_{11}/2)\int ({\bf 
\nabla \cdot N})^2 dV+(2k_{11}+k_5)(\rho_o-\rho_i)\oint KdA_i=k_c\int
\kappa (s)^2 ds+\pi k_{11}\ln (\rho_o/\rho_i)\int ds$, here $k_{ij}$ is the
Oseen-Frank elastic constants and $k_5=2k_{13}-k_{22}-k_{24}$, $k_c=\pi
k_{11}(\rho _o^2-\rho _i^2)/4$, and ${\bf N}$ and $K$ are respectively the
director and Gaussian curvature defined on the inner surface of the Sm-A
tube; $\kappa $ is the curvature defined along ${\bf r}$ [14, 15].

 The total energy
difference is thus $F=F_V+F_A+F_c=k_c\int \kappa(s)^2+\int (\lambda +\sqcap )ds$,
 where $\lambda =\pi k_{11}\ln (\rho_o/\rho_i)+2\pi \gamma (\rho_o+\rho_i)$ 
 and $\sqcap =-\pi g_0(\rho_o^2-\rho_i^2)$ [14]. We call $F$ the {\it shape
formation energy} of the Sm-A filament. The variation equation $\delta F=0$
yields the equilibrium-shape equation of the filament: 
\begin{eqnarray}
k_c(\kappa ^3-2\kappa \tau ^2+2{\ddot{\kappa}})-(\lambda +\sqcap )\kappa =0,\\
4{\dot{\kappa}}\tau +2\kappa {\dot{\tau}}=0,
\end{eqnarray}
where $\tau$ is the torsion of the filament's central line ${\bf r}$,
according to the Frenet formula [16]; ${\dot{(\; )}}$ and ${\ddot{(\; )}}$
mean respectively first  and second order differentiations with respect to
arc length $s$.

It is obvious that a straight line is always a solution of the above
equation, since  its $\kappa=\tau=0$. The corresponding
shape formation energy of a straight filament is $F=(\lambda +\sqcap )l$,
where $l$ is the length of the straight filament. In the rapid cooling
process, it is impossible to effectively transfer redundant heat out of the
system by means of heat exchange with environment,
 and Sm-A straight filament will form
spontaneously and quickly to partly accommodate the redundant energy, {\it
 as
long as} $\lambda +\sqcap >0$. Straight Sm-A filaments have been observed to
form at the initial stage of the growth process [14]. However, since the
 value of
$g_0$ increases with temperature decreasing [14, 17], while $\gamma $ exhibits no
temperature dependence or a weak one [18], $\lambda +\sqcap $ will decrease with
cooling, and the equilibrium threshold condition of $F=0$ yields the
criteria for the growth of a straight filament as 
\begin{equation}
\pi k_{11}\ln (\rho _o/\rho _i)+2\pi \gamma (\rho _o+\rho _i)-\pi g_0(\rho
_o^2-\rho _i^2)=0,
\end{equation}
beyond which, forming a straight filament will no longer help to absorb
thermal energy.  This situation is unfavorable to the quench process, and
consequently
the resultant remnant part of energy will prevent the straight filament from
keeping stable. Then a shape deformation will be induced, which would
lead to another solution of the shape equation to convert as much
amount of thermal energy
into elastic deformation energy as possible. In other words, the
straight filament will buckle to a curved configuration which, {\it on the one
hand, its} $\kappa $ {\it and} $\tau $ {\it should satisfy} (1) {\it
and} (2) {\it so the new
pattern will be stable, and on the other hand, the stable new configuration
should be capable of absorbing as much thermal energy as possible}. Then,
to what configurations will the filament deform?

Now we would prove that the observed Sm-A serpentine planar patterns shown
in Ref.~[14] are just the allowed solutions of Eqs.~(1) and (2).  For the convenience of comparison with experimentally observed
planar curved patterns, we will confine ourselves to the  planar
case, where $\tau =0$. Then Eq.~(2) is automatically satisfied and (1)
reduces to 
 \begin{equation}
k_c(\kappa ^4+2\kappa {\ddot{\kappa}})-(\lambda +\sqcap )\kappa ^2=0,
\end{equation}
here, we have  multiplied both sides of Eq.~(1) with $\kappa $.
Define the filament central line tangent unit vector to be ${\bf t}(s)=(\cos
\phi (s),\sin \phi (s),0)$, then $\kappa ^2={\dot{\phi}}^2$. Inserting this
into (4), we know that $\phi $ is determined by  the following equation: 
\begin{equation}
{\dot{\phi}}^2=-2\eta +{\frac f{k_c}}\cos (\phi -\phi _0),
\end{equation}
where $\eta =-(\lambda +\sqcap )/2k_c$ [14] and $f$ is a positive
 constant, (later, we will show that $f$ is actually the mean repulsive force exerted on
the growing end of the filament), $\phi_0$ is an integration constant. 
The  solution of (5) is a periodic
function, its form in the range $\phi \in (0,\arccos (1-2m))$ is expressed
as [14]
\begin{equation}
\sqrt{{\frac{2k_c}f}}F(\arcsin \sqrt{{\frac{1-\cos \phi }{2m}}},\sqrt{m})=(s-s_0),
\end{equation}
here, 
\begin{equation}
m={\frac 12}-{\frac{k_c\eta }f}\;\in\;(0,\;1)
\end{equation}
and $s_0$ is another integration constant, $F(\psi ,k)$ is the first kind elliptic
integral. In the same range the filament shape is determined, up to an
additive constant vector, by [19]
\begin{equation}
{\bf r}(s)=\sqrt{{\frac{2k_c}f}}\left( 2E(\arcsin \sqrt{{\frac{1-\cos \phi }{2m}}},
\sqrt{m})-F(\arcsin \sqrt{{\frac{1-\cos \phi }{2m}}},\sqrt{m}),-\sqrt{2m+
\cos \phi -1}\right) 
\end{equation}
where $E(\psi ,k)$ is the second kind elliptic integral. The trajectory of
the filament in the whole range can also be obtained easily from (5), (6)
and (8). The general shapes are shown in Fig.~1 in the cases of $m<1/2$ or $\eta >0$ (dotted curve), $m=1/2$ or $\eta =0$ (dashed curve) and $m>1/2$ or $\eta <0$ (solid curve) for
one period of the trajectory.
The total arc length of each period is calculated to be  [19]
\begin{equation}
P=4\sqrt{{\frac{2k_c}f}}K(\sqrt{m})
\end{equation}
and the shape formation energy density of each period is [19]
\begin{equation}
\chi ={\frac 1P}\int_0^P(k_c{\dot{\phi}}^2-2\eta )ds=f\left[
2L(\sqrt{m})-3+4m\right] ,
\end{equation}
where $K(k)$ and $E(k)$ are respectively the first and second kind complete
elliptic integrals, and $L(\sqrt{m})=E(\sqrt{m})/K(\sqrt{m})$.

At the initial stage of buckling, the filament will deform to a  configuration
to make the threshold condition $\chi =0$ be satisfied, which leads to
 $m=m_i=0.339175$ at
this stage. The corresponding deformed shape is the dotted curve of Fig.~1, it
is very similar to the observed initial buckling shape Fig.~1b and 1c of 
Ref.~[14]. 
Furthermore, for this shape, the amplitude of the
deformation is calculated to be $Y_{max}=\sqrt{4mk_c/f}$ from (8), and the
distance traversed along the growth direction  for one half of the 
trajectory period is $X_{max}=\sqrt{2k_c/f}(2E(\sqrt{m})+K(\sqrt{m})$, thus $Y_{max}/X_{max}=0.737$ for the above mentioned value of $m_i$. On the
experimental side, we can estimate the values of this ratio to be respectively about $0.85$ and $0.64$ from Fig.1c
and 1d of Ref.~[14], and the mean value is $0.745$, all
 in close agreement with the theoretical value.

As mentioned before, after buckling, the filament will try its best in
adjusting its structures to accommodate thermal energy. 
Differentiation of $\chi $ in (10)
with respect to $m$, we find that the maximum value of the shape formation
energy density $\chi $ is reached at the point where $m_o=0.9689$, and in
the range between $m_i$ and $m_o$, $\chi $ is a continuously increasing
function of $m$. Thus, after the buckling process has been triggered, $m$
will try to increase from its initial value $m_i$, till the value $m_o$ is reached.
 Since $m$ is determined
by (7), we anticipated that $\lambda +\sqcap $ will  change from
negative value ( and hence $m<1/2$) to zero ($m=1/2$) and then to become more and more
positive ($m>1/2$). The question is: How can this be achieved by the
Sm-A filament? 

We know that $g_0$ increases with temperature decreasing,
 and $\gamma$ exhibits little temperature dependence, and that during the
 cooling process there is no significant changes in the filament radii $\rho_o$
and $\rho_i$, therefore, the only possible reason accounting for $\lambda+\sqcap$
changing from negative to positive values is that $k_{11}$, the quantity
related to the molecular structures of Sm-A filaments, rapidly increases its value.
In other words, quench may induce some 
structural transition in the Sm-A filament at
the molecular level. (In Ref.~[14] it is speculated that such a transition is
the {\it trans} to {\it cis} transition occurred in the  alkyl chains of the
Sm-A liquid crystal material. Further experimental observations are still
needed in this respect).
 Consequently, the initial curved pattern  will take the
form demonstrated by the dotted curve in Fig.~1,  
later formed segments will take the form of  the 
dashed curve, and still later formed segments will be the form
  of the solid curve.
 This trend is demonstrated more clearly  in Fig.~2. Buckled pattern resembling that of
 Fig.~2 has
 indeed been  observed and 
recorded in the experiment
of Ref.~[14] (see Fig.~1 of [14], especially 1d), showing the
correctness of the pattern formation mechanism based on competition between heat transfer and
heat conversion.

When $m$ is increased to the value $m_c=0.731183<m_o$, different segments of
the same filament begin to intersect, as shown in Fig.~3.
  At this stage, the formed
serpentine patterns may begin to collapse because of possible strong
contacting forces,  and eventually the filament will 
transform to compact domains [14]. Thus actually
the optimal value $m_o$ can not be reached by the system.

So far, we have not discussed the physical meaning of the parameter $f$ in
Eq.~(5). Now we begin to investigate into this question. Our result is that
 $f$ is actually the mean repulsive force exerted on the filament
growing end. There has been clear experimental evidence that both the
assembly and disassembly of microtubular filaments can generate force;
however, only limited quantitative data are available on the actual
magnitude of these forces [20]. In the case of Sm-A filaments discussed here, we
estimate the force exerted on the ends to be the order of $10^{-1}$
piconewton. This magnitude is reasonable as indicated by various experiments
on thin filaments [20, 21]. The forces associated with the assembly and
 disassembly of
filaments may be of vital biological significance. For example, it has long
been speculated that the motion of chromosomes during mitosis of the cell
cycle is caused by the assembly and disassembly of
 cytoskeletal microtubes [20].

To show that the parameter $f$ in Eq. (5) corresponds to the average force
exerted at the ends, we will rederive (5), the stable planar shape equation,
from another way. The already grown Sm-A filament can be viewed as an
 inextensible
string. When subjected with the constraint of fixed end-to-end distance, the
total shape energy is expressed as $F=\int \left[k_c{\dot{\bf t}}^2+\lambda
+\sqcap -{\bf f\cdot t}\right]ds$, with ${\bf t}(s)=(\sin \theta \cos \phi ,\sin
\theta \sin \phi ,\cos \theta )$ being the string direction vector; here
${\bf f}$ is a Lagrange multiplier corresponding to this constraint, which, 
according to thermodynamic principles,  is just the average force exerted
by the environment on the filament ends. The first variation $\delta F=0$
 leads to the shape equation expressed
in terms of $\theta $, $\phi $ and their differentiations with respect to $s$,
 For our planar case with $\theta =\pi /2$, the non-straight stable
configurations are found to be determined by the following equation 
\begin{equation}
{\dot{\phi}}^2=C+{\frac{|{\bf f}|}{k_c}}\cos (\phi -\phi _0^{\prime }+\pi ),
\end{equation}
where $\phi _0^{\prime }$ is the angle between the x-axis and the direction
of the average force ${\bf f}$. From (11) we know the filament tangent
direction (growth direction) ${\bf t}$ is on  average antiparallel to 
${\bf f}$, therefore ${\bf f}$ must be a repulsive force. Comparing Eq.~(11)
 with (5), we can obviously infer that the parameter $f$ in (5) is
just the magnitude of the average  force ${\bf f}$ and $C=-2\eta $
 in (11).
This force is induced by thermal current, and we have assumed it to be
constant throughout the growing process.

The magnitude of the force can also be estimated from the buckling
amplitude of the filament. As mentioned before,  the deformation amplitude $%
Y_{max}=\sqrt{4mk_c/f}$, therefore  $f=4mk_c/Y_{max}^2$. At the initial
stage of buckling, $m=m_i=0.339175$, we take $k_{11}=10^{-6}$ dyn according to
Ref.~[13], and
estimate that $\rho _o=1.3$ $\mu $m, $\rho _i=0.8$ $\mu $m, $Y_{max}=10.4$ $%
\mu $m from fig.1c of Ref.~[14]. Thus, the average repulsive force is
calculated to be $f\simeq 1.4\times 10^{-13}$N or $0.14$ pN. It is
interesting to note that the forces related to polymeric strings are often
in  the order of $^{}10^{-2}$ to $10^{1}$ pN [20, 21].

In summary, in this paper we have proposed a theory to explain the novel
pattern formation phenomena induced by catastrophic variations of environmental
temperature or other parameters. We then applied this general insight in
understanding the pattern formation process of smectic-A filaments grown from
an isotropic liquid phase. Exact buckling patterns have been obtained and
compared with experiment, and we achieved excellent agreements with experiment. 
 We also suggested that temperature decreasing leads to some structural
transitions at the molecular level for the Sm-A filament, possibly the
{\it trans} to {\it cis} transition of the alkyl chains of the material used in
the experiment of [14]. The force associated with the pattern formation
process of smectic-A filaments is estimated to be in the order of $0.1$ pN.

One of the author (Z.H.)  appreciates valuable   discussions with
Drs. Yan Jie, Liu Quanhui and Zhao Wei. We are also
indebted to the anonymous
Referee, who makes us take notice of Refs.~[6] and [7].
 This work is partly supported by
the National Natural Science Foundation of China.

\begin{figure}
\caption{Shapes of one  period  of the buckled patterns (Eq.~(8)) for 
$m=0.339175$ (dotted line), $m=0.5$ (dashed line) and $m=0.7$
 (solid line). The length unit  is set to be $\sqrt{2 k_c/f}$.}
\label{FIG. 1}
\end{figure}

\vskip 0.2in
\begin{figure}
\caption{Simulated buckling filament shape  during the 
growth process.
Later formed  segments have longer period arc
length and larger amplitude, caused by increased values of $k_{11}$.
The unit length is set to be $\sqrt{2 k_{c}^i/f}$, where $k_{c}^i$ is 
the value of $k_c$ at the onset of buckling.}
\label{FIG.2}
\end{figure}

\vskip 0.2in
\begin{figure}
\caption{Onset of self-intersection of the buckled pattern at $m=0.731183$. The 
appearance of self-intersections might 
trigger the transition from filament structures to compact domains. The
length unit is the same as in Fig.~1.}
\label{FIG. 3}
\end{figure}
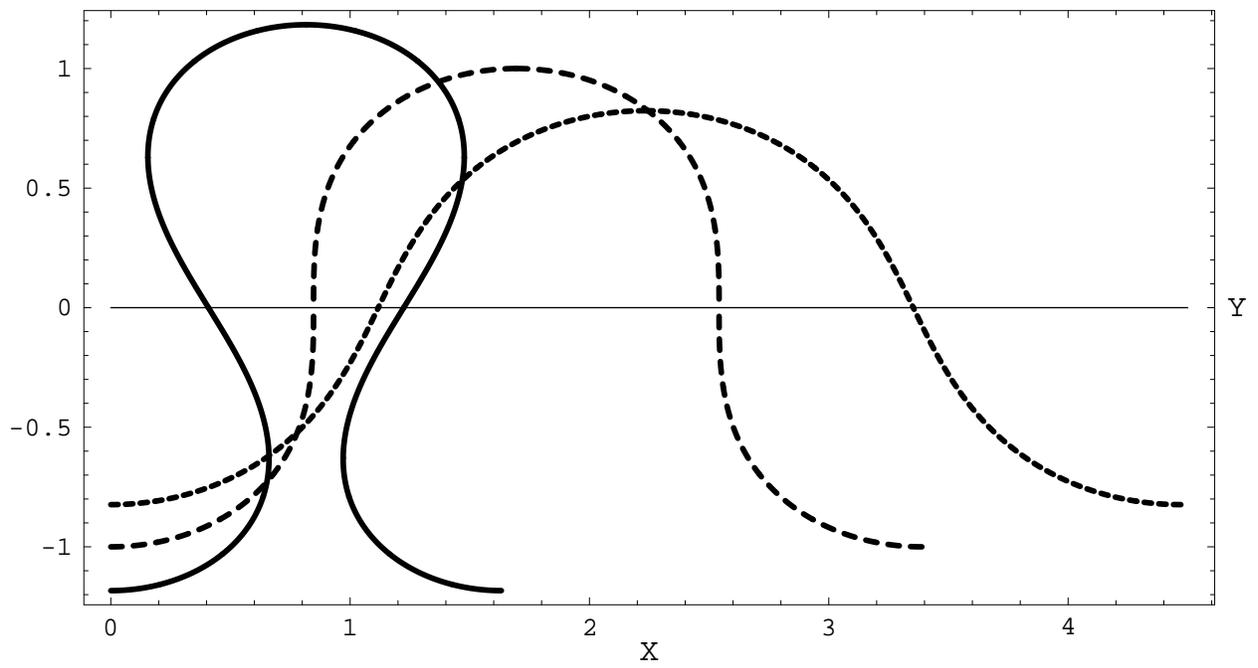
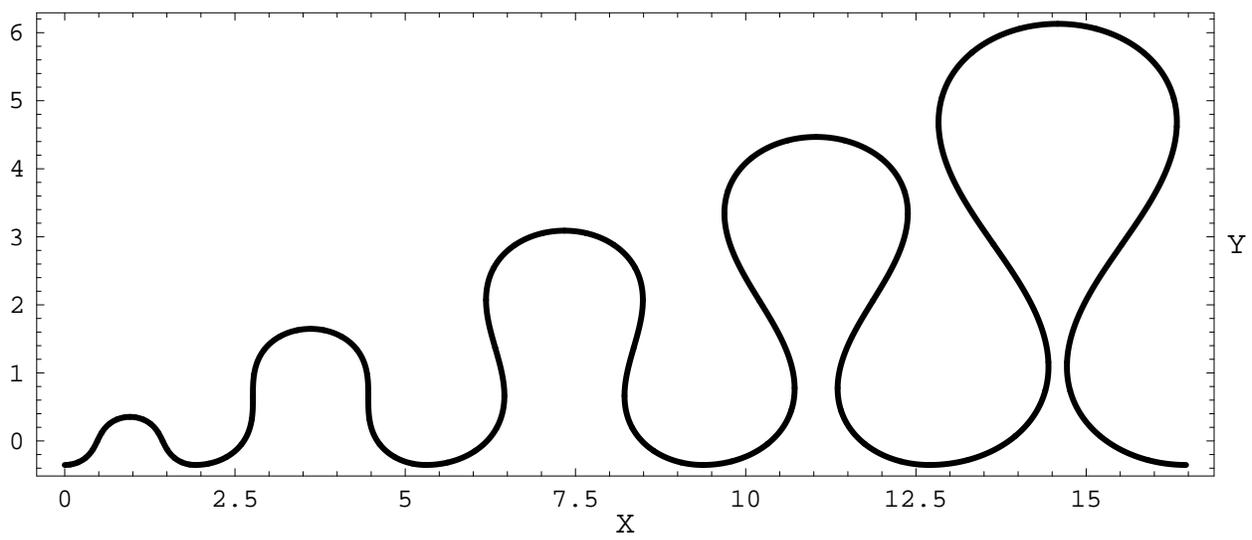
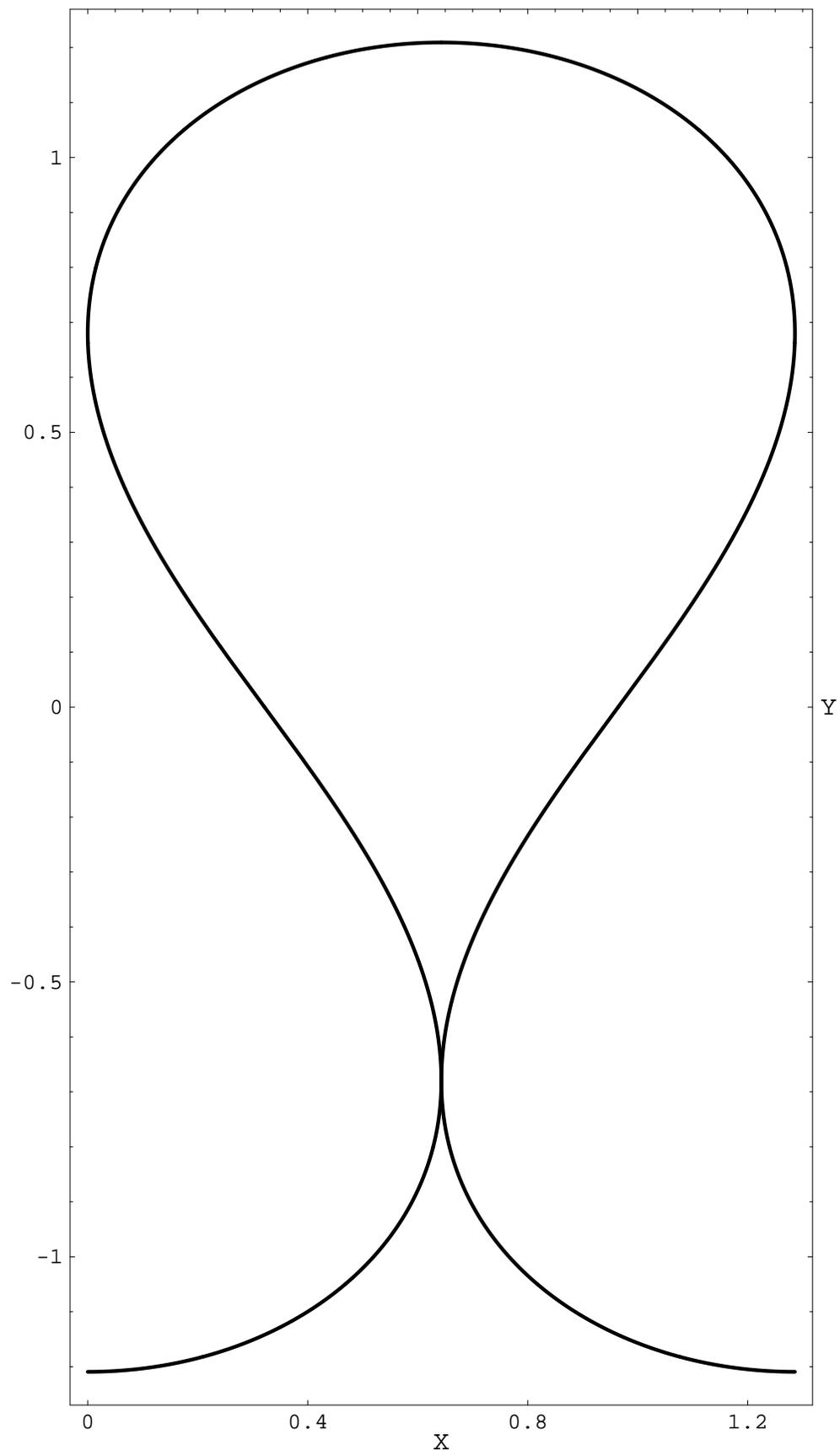
\end{document}